\documentclass[twocolumn,showpacs,preprintnumbers,amsmath,amssymb]{revtex4}
\input psfig.sty
\usepackage{graphicx}
\usepackage{dcolumn}
\usepackage{bm}
\usepackage{latexsym}
\usepackage{epsfig}
\begin{document}
 \newcommand{\bq}{\begin{equation}}
 \newcommand{\eq}{\end{equation}}
 \newcommand{\bqn}{\begin{eqnarray}}
 \newcommand{\eqn}{\end{eqnarray}}
 \newcommand{\nb}{\nonumber}
 \newcommand{\lb}{\label}
\title{Black Hole Formation with an Interacting Vacuum Energy Density}
\author{M. Campos $^{1,2}$} \email{m.campos@usp.br} 
\author{J. A. S. Lima $^1$} \email{limajas@astro.iag.usp.br}

\vskip 0.5cm
\affiliation{$^1$Departamento de Astronomia, Universidade de S\~ao Paulo, 05508-900 S\~ao
Paulo, SP, Brasil}
\affiliation{$^2$Departamento de F\' \i sica , Universidade Federal de Roraima, 69304-000 Boa Vista, RR, Brasil.}
%
%
\date{\today }
\begin{abstract}
We discuss the  gravitational collapse of a spherically symmetric massive core of a star in  which the fluid component is interacting with a growing vacuum energy density.  
The influence of the variable vacuum in the collapsing core is quantified by a phenomenological  
$\beta$ parameter as predicted by dimensional arguments and the renormalization group approach. 
For all reasonable values of this free  parameter, we find that the vacuum energy density increases the collapsing time but it cannot prevent 
the formation of a singular point. However, the nature of the singularity depends on the values of $\beta$. In the radiation case, a trapped surface is formed for $\beta \leq 1/2$ whereas for $\beta \geq 1/2$, a naked singularity is developed. In general, the critical value  is $\beta= 1- 2/3(1+\omega)$, where $\omega$ is the parameter describing the equation of state of the fluid component.   
\end{abstract}
\vspace{.7cm}
\pacs{97.60.-s, 95.35.+d, 97.60.Lf, 98.80.Cq}
\maketitle
%
\section{Introduction}
%
\renewcommand{\theequation}{1.\arabic{equation}}
\setcounter{equation}{0}

It is widely known that the observed Universe is undergoing an expanding accelerating stage \cite{Riess, Komatsu}. 
The simplest explanation in the context of Einstein's general relativistic theory (GRT)  is the existence of a new dark component (in addition to cold dark matter)  
whose energy density remains constant or slowly varying in the spacetime. The
most theoretically appealing possibility for the so-called dark energy is
the energy density stored on the true vacuum state of all
existing fields in the Universe, i.e., $\rho_{v} = \Lambda_0/8\pi G$, where
$\Lambda_0$ is the cosmological constant. At the level of GRT, the $\Lambda$-term is usually interpreted  as a relativistic simple fluid with equation of state (EoS), $p_v = - \rho_v$ \cite{rev1,Bas2010}.

The so-called cosmic concordance model ($\Lambda$CDM),  a flat  cosmology with 
baryons, cold dark matter plus a relic $\Lambda$-term,
seems to be in agreement with all cosmological
observations available. From the theoretical viewpoint,
however, the well-known cosmological constant problem,
i.e., the unsettled situation in the particle physics/cosmology
interface in which the cosmological upper bound
($\rho_{v} \lesssim 10^{47} GeV^4$) differs from theoretical expectations
($\rho_{v} \sim 10^{71} GeV^4$) by more than 100 orders of magnitude,
originates an extreme fine-tuning problem \cite{Weinberg}. 

A natural attempt of alleviating the so-called cosmological constant problem is 
to allow a time dependence of $\Lambda$ or equivalently, of the vacuum energy density. 
Historically, the idea of a time varying $\Lambda(t)$-term was first advanced
in the paper of Bronstein \cite{B33}. Different from Einstein's cosmological
constant, such a possibility was somewhat missed in
the literature for many decades, and, probably, it was not
important to the recent development initiated by Ozer and
Taha \cite{OT86} at the late eighties. After their papers, a number of models with different
decay laws for $\Lambda(t)$ 
were proposed by many authors and their predictions  confronted with the available 
observational data \cite{L1,CLW,L2,L3,L4}. It is worth mentioning
that the most usual critique to these $\Lambda(t)$CDM scenarios
is that in order to establish a model and study their
observational and theoretical predictions, one needs first
to specify a phenomenological time-dependence for $\Lambda$. However, there are some attempts to represent 
out of equilibrium dynamical $\Lambda$ 
models by a  scalar field \cite{ML02,EA1}, as well as 
based on a Lagrangian description \cite{Lag1}.  

Besides the evolution of the Universe, an important process in gravitational 
physics that can also be affected by a dynamical $\Lambda (t)$-term is the formation of black holes. 
A basic difference here is that the 
interacting vacuum energy density decreases in the expanding Universe whereas 
during the black hole formation (a collapsing process), it grows in the course of time.  
Therefore, it is natural to ask about the influence of a  
growing vacuum energy density during the gravitational collapsing process of a  
star. In principle, since a time varying $\Lambda (t)$ exerts an increasing repulsive force
on its surrounding medium, it might also prevent the ultimate formation of a spacetime singularity.

Another closely related issue is the possible influence of $\Lambda(t)$  on the cosmic censorship hypothesis (CCH), as well 
as on the nature of the singularity.  
In its weak form, this conjecture eliminates the occurrence of naked singularities in the spherical gravitational 
collapse whereas its strong version states that all singularities in any realistic spacetime are never 
visible to a distant observer because are hidden behind an event horizon \cite{Penrose}. Since the 
earlier counter example to the CCH discussed by Papapetrou \cite {Papa}, the emergence of naked  
singularities or black holes has been intensively investigated in the  
literature, including the effect of different material components \cite{Collapse}. However, as far as 
we know, the possible influence of a time varying $\Lambda$-term in the last stages of a collapsing 
system (including the formation of a trapped surface and naked singularities) has not been analyzed in the literature. In principle, this is an important issue due two 
combined effects: (i) unlike what happens in an expanding Universe, the energy density of a coupled vacuum component 
grows in the course of the gravitational contraction, and (ii) since the vacuum pressure is negative  
and  generates repulse gravity, potentially, it might alter significantly the late stages of any collapsing matter distribution. 

In this paper, we discuss the formation of black holes (and naked singularities) during  
the gravitational collapse  of a fluid interacting with a time-varying vacuum  
energy density. For given initial conditions, the equations describing  the evolution of the two-fluid interacting 
mixture are analytically solved, but, in order to study the
different roles played by the matter equation of state during the collapse of the core, a special attention  
is dedicated to the dust and radiation cases. 
The formation of the black holes here is simply identified with the development of
apparent horizons before the formation of the singularity. As we shall see, due to its repulse 
gravitation, a time-varying vacuum energy density as modeled here increases the collapsing time 
but under certain conditions it cannot prevent the formation of black holes. Our result also 
suggest that the CCH conjecture (at least in its weak form) is generically  violated in the presence 
of a time varying vacuum due to the formation of naked singularities. 
%
\section{Collapsing spherical star with a growing vacuum component}
%
\renewcommand{\theequation}{2.\arabic{equation}}
\setcounter{equation}{0}
\subsection{Star Medium: Composition and Geometry}
Let us now consider the gravitational collapse of a spherically
symmetric massive core of a star with finite thickness. 
The massive core medium is formed by a two fluid interacting mixture: a material component 
plus a growing vacuum energy density represented by a $\Lambda(t)$-term.  

To begin with,  let us  divide the spacetime into three different regions, $\Sigma$ and $V^{\pm}$, where $\Sigma$ denotes  
the surface of the star,  and $V^{-}\;$ ($V^{+}$) the interior (exterior) 
of the massive core. For the sake of simplicity, we also assume that the 
spacetime inside the massive core is homogeneous and isotropic, a particular case of the 
inhomogeneous Oppenheimer-Snyder model \cite{Oppenheimer}.  This means that the spacetime inside the core
is described by the homogeneous and isotropic flat Friedmann-Robertson-Walker (FRW) geometry: 
\bq
\lb{2.1}
ds^{2}_{-} = dt^{2} - a^{2}\left(t \right)\left(dr^{2} + r^{2} d\Omega^{2}\right), 
\eq
where  $a\left(t \right)$ is the scale factor and $d\Omega^{2} \equiv d\theta^{2} + \sin^{2}\theta d\varphi^{2}$ is the area element on the unit sphere.
Although, this is a very ideal case, we do believe that it captures the main features of 
gravitational collapse in the presence of a growing vacuum energy density. We recall that  for expanding Universe models the curvature effects are not important at the early stages of the evolution \cite{LL}. Similarly, it will be assumed here that the same happens for the late stages of the collapsing core.  

Following standard lines, in this paper we shall focus our attention mainly in the spacetime 
inside the star \cite{CaiWang}.  If the collapsing 
massive core finally forms a black hole (BH), an apparent horizon {\em must} 
develop inside it, and, therefore, there  exists a moment 
at which the whole core collapses inside the apparent horizon.    
 
The apparent horizon is a trapped surface lying in a boundary of a particular surface $S$, and can differ from the intersection of the event horizon 
with the surface $S$, where the event horizon is the boundary of the region $S$ that is not possible to scape to infinity \cite{Hawking1}.
Using a more technical language, consider a two-sphere $S$ embedded in a slice $\Sigma$ of the spacetime $M$, and let $s^ \mu$ be the
 outward pointing spacelike unit normal to the $\Sigma$, and $n^ \mu$ the future-pointing timelike unity normal to $\Sigma$.  
Hence, the vector $\kappa ^\mu = s^\mu +n^\mu$ is a nulll vector, and $S$ is a marginally trapped surface if $\kappa ^\mu _{;\mu} = 0$ 
holds everywhere on the $S$ \cite{Anninos}.

Although, considering that the matching conditions 
and the spacetime outside the star must affect the total mass and the global structure of the 
black hole,  the key aspect of black hole formation is to know whether apparent horizons develop
inside the core. In other words, the ultimate formation of the black hole singularity does not depend neither on the matching nor the choice of the 
spacetime outside the star. Here we are mainly interested to discuss under which conditions  a BH is formed during the collapse of  
dust and radiation fluids when such components are interacting with a growing vacuum energy density. 

Let us first consider the Einstein field equations within the star:
\begin{equation} \label{EE}
G_{-}^{\mu \nu} = \chi \left[T_{-}^{\mu \nu} +
\frac{\Lambda} {\chi}g_{-}^{\mu \nu}\right],
\end{equation}
where $G_{-}^{\mu \nu}$ is the Einstein tensor  and  $T_{-}^{\mu \nu}$ is the energy-momentum
tensor, and $\chi =  8 \pi G$
($c = 1$) is the Einstein's constant. According to 
Bianchi identities, the above equations imply that the $\Lambda$ is constant 
only  if  $T_{-}^{\mu \nu} \equiv 0$ or separately conserved, i.e., $T_{-}^{\mu \nu};_{\nu} =
0$. This means that a time varying vacuum
is possible only by assuming the previous existence of some sort of
non-vanishing interacting fluid which is changing energy with the vacuum component. 

{ Before to  proceed further,  it should be  stressed that the basic discussion here is related to black holes and naked singularities formed from collapsing star cores. In particular,  this means that supermassive black holes like the ones found in the galactic centers are not part of our investigation, and the same happens with the recently discovered  quasar  at redshift $z=7.085$ and estimated mass $M=2.10^{9}M_{\odot}$ \cite{Mortlock}. When such object was formed,  the Universe  had less than one billion 
of years after the big-bang. Therefore, it will be assumed here that such supermassive structures were seeded by massive collapsing star cores  of population III,  but their extremely large mass  is the result of cosmological accretion mechanism and mergers in the course of the evolution.}

\subsection{Basic Equations and Solutions.}

In what follows, it will be also explicitly assumed that the vacuum and fluid components are coupled and that $T_{-}^{\mu \nu}$ has the form of a perfect fluid:

\begin{equation}
T_{-}^{\mu \nu}= \left(\rho_f + p_f\right)u^{\mu}u^{\nu}
- p_f g_{-}^{\mu \nu},
\end{equation}
where $\rho _f$, $p _f$ and $u^{\mu}$, are the energy density, the pressure and  the 4-velocity of the fluid component, respectively.  

In this case, by taking the divergence of ({\ref{EE}}) and projecting the result in the direction of the four-velocity one finds:
\begin{equation}
\label{coupling}  u_{\mu}{T_{-}^{\mu \nu}}_{;\nu} =
-u_{\mu}\left(\frac{\Lambda g^{\mu \nu}_-}{\chi}\right)_{;\nu}. 
\end{equation}

\begin{figure*}[th]
                 \centerline{\psfig{figure=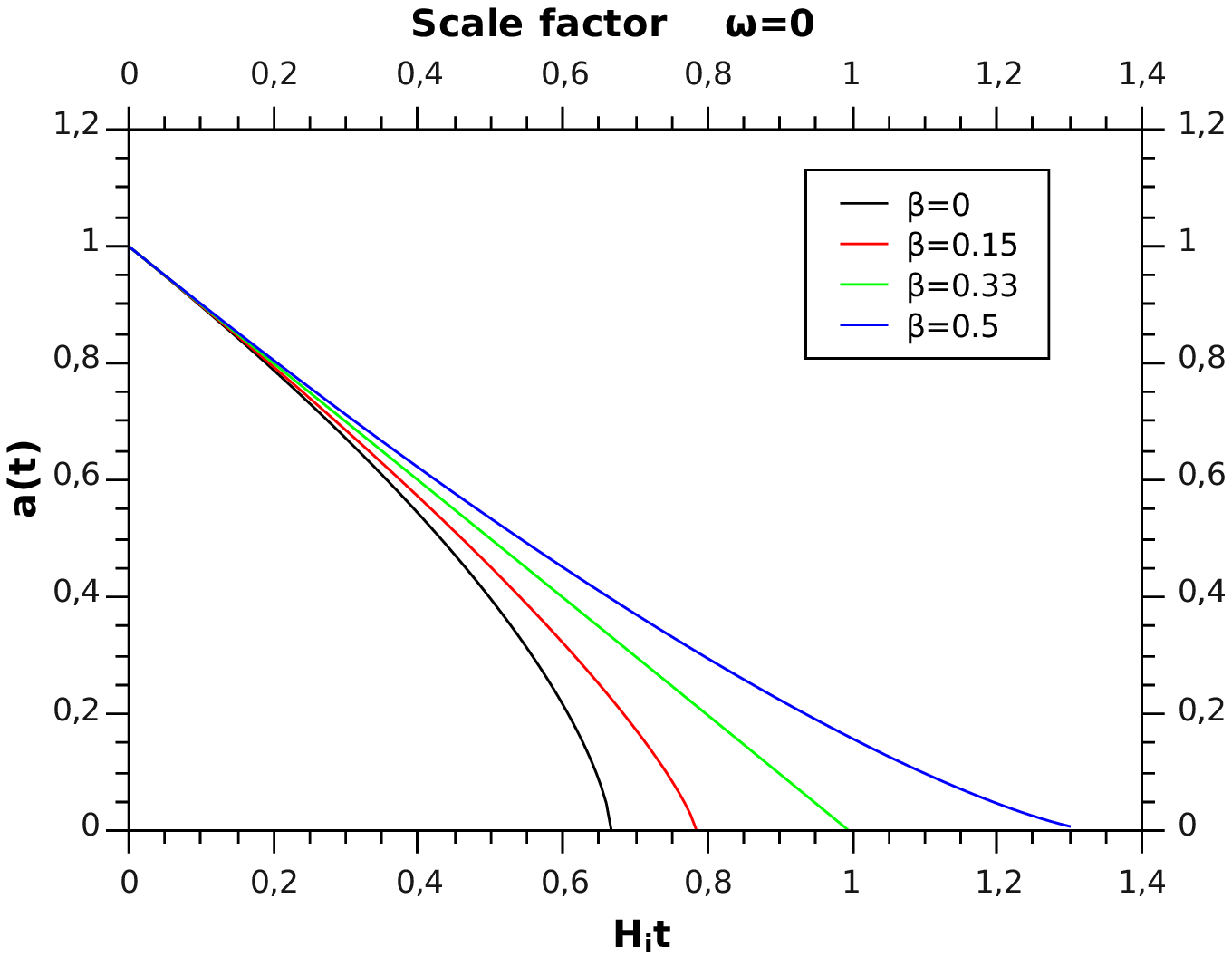,width=3.0truein,height=1.9truein}
                    \psfig{figure=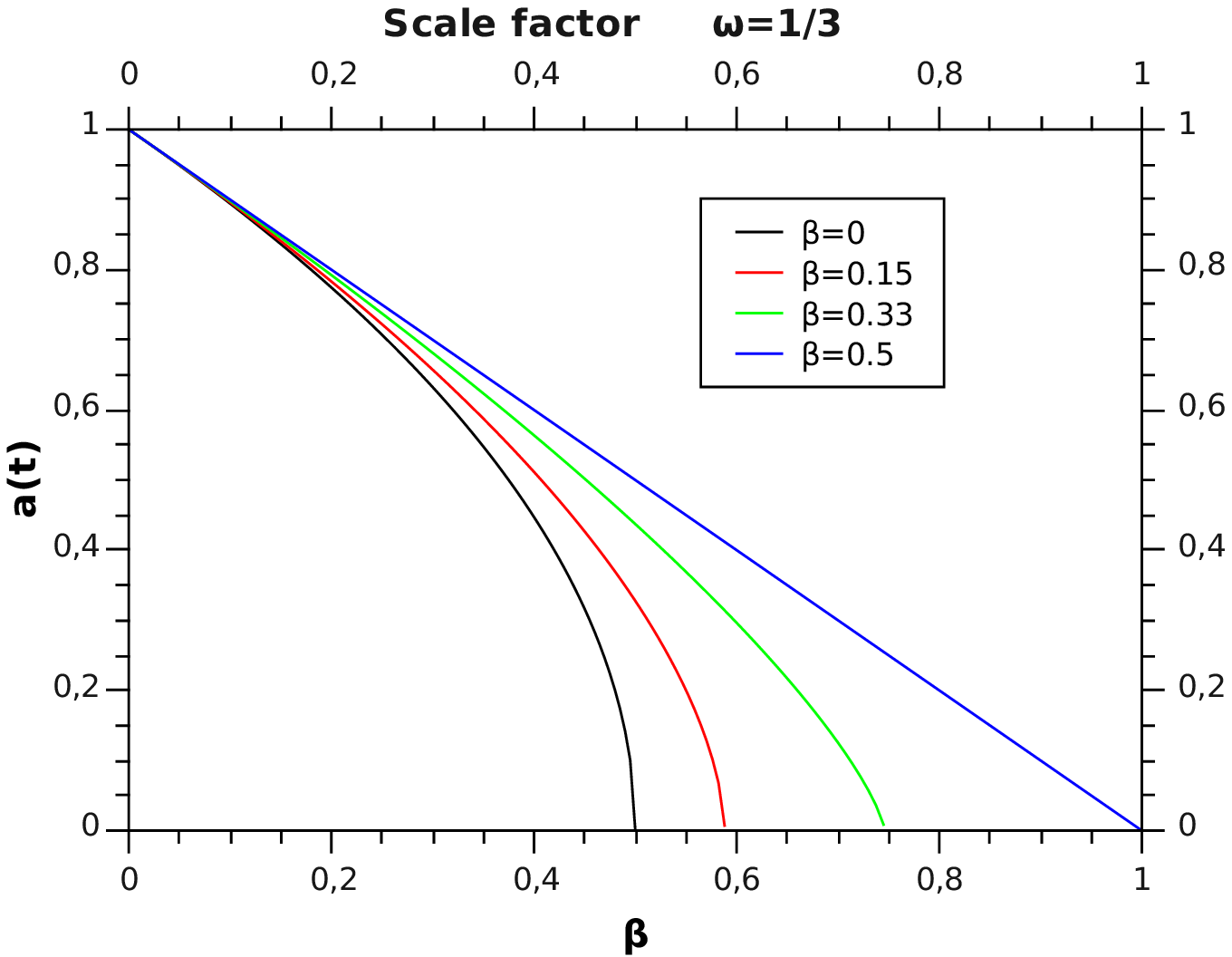,width=3.0truein,height=1.9truein}}
                      \caption{Evolution of the scale factor for the collapsing mixture. The right and left panels correspond to dust ($\omega=0$) and radiation ($\omega=1/3$) coupled to  a growing vacuum component, respectively. The overall effect of the vacuum component is to increase the time of collapse with respect to a pure fluid medium. However, the total collapse is not avoided for the selected values of $\beta$ depicted in the panels. Similarly, the same happens for any value of $\beta < 1$ (cf. Eq. (2.12) and comments in the body of the text).}
                     \label{A}
                    \end{figure*} 

In the background (\ref{2.1}), the above energy conservation law for the comoving observer reads: 

\begin{equation}\label{coupling1}
\dot \rho_f + 3\frac{\dot a}{a}\left(\rho_f + p_f\right) \ = - \dot \rho_v,
\end{equation}
where a dot means time derivative and $\rho_v = \Lambda(t)/8\pi G$ is the time-dependent vacuum energy density. 

As in the standard case (without a $\Lambda(t)$-term), the energy conservation law (\ref{coupling1}) is also contained in the Einstein field equations:

\begin{eqnarray}
8\pi G \rho_f +\Lambda \left(t \right) &=&3H^2, \\
8\pi G p_f -\Lambda \left( t \right ) &=&-2\dot{H}-3H^{2},
\end{eqnarray}
where $H=\dot a/a < 0$ is the ``Hubble function'' for the collapsing mixture. 
               
In order to solve the above equations we need to specify the fluid equation of state and the time varying vacuum energy density. 
For the sake of definiteness, it will be assumed that the matter component of the star satisfies a barotropic equation of state (EoS)
\begin{equation}\label{EoS}
p_f=\omega \rho_f \,, 
\end{equation}
where  $0\leq\omega\leq 1$ is a positive parameter (here we are not particularly interested in the case of a pure or interacting dark energy fluid). 
Note that if the collapsing fluid is itself a mixture (for instance, matter and radiation),  we are implicitly assuming that the variable 
vacuum is interacting only with the dominant fluid component, and that such a mixture determines the overall evolution 
of the collapsing medium.

Many phenomenological functional forms have been proposed in the literature for describing a time-varying $\Lambda(t)$. Based on dimensional arguments, 
Carvalho {\it et al.} \cite{CLW} shown that a natural dependence is $\Lambda \propto H^{2}$. Later on, this functional
 dependence  was derived within a renormalization 
group approach (including a bare cosmological constant $\Lambda_0$) by Sol\`a and Shapiro \cite{SS}. Following these authors,  
we consider here that the  $\Lambda(t)$-term  is given by \cite{CLW,SS}: 
\begin{equation}\label{Lambda}
\Lambda = \Lambda _0 +3\beta H^{2},
\end{equation}
where $\beta$ is a dimensionless constant parameter and the factor 3 was added for mathematical convenience.  

Using expressions (\ref{EoS}) and (\ref{Lambda}), a simple manipulation shows that the scale factor satisfies: 
\begin{equation}
\frac{\ddot a}{a} + \left[ \frac{3}{2}(1+\omega)(1-\beta)-1 \right]\frac{\dot a ^2}{a ^2} -(1+\omega)\frac{\Lambda_0}{2} = 0.
\end{equation}
Now, by integrating the above equation we obtain: 
\begin{equation}\label{a(t)}
a(t)=a_i \left[(\frac{1-\tilde{\Omega}_{\Lambda_{0i}}}{\tilde{\Omega}_{\Lambda_{0i}}})^{\frac{1}{2}} {\sinh \xi {\sqrt{\tilde{\Omega}_{\Lambda_{0i}}}}(t_c - t)}   
 \right] ^{\frac{2}{3(1+\omega)(1-\beta)}},
\end{equation}
where $\xi = \frac{3}{2}(1-\beta)(1+\omega )H_i$, $\tilde{\Omega}_{\Lambda_{0i}}={\Omega_{\Lambda_{0i}}}/{(1-\beta)}$,  and the positive quantities, $a_i$, $H_i$, 
define the initial conditions of the collapsing core [$a(0)=a_i$, $H(0) = - H_i$]. We have also defined the initial
 vacuum (bare) density parameter by $\Omega_{\Lambda_{0i}} = \Lambda _{0}/{3 H_i ^2}$  whereas the collapsing time, $t_c$,
 is fully determined by the initial condition $a(0) = a_i$, and, as such, it depends only on the physically meaningful
 parameters ($H_i$, $\beta$, $\omega$ and $\Omega_{\Lambda_{0i}}$). 

The main aim here is to understand the influence of the growing vacuum energy density on the final stages of the collapsing  
process. In this way, one may conclude that the contribution of the bare cosmological constant $\Lambda_0$ becomes rapidly negligible 
in comparison to the variable $\Lambda(t)$-term  ($\propto H^{2}$). Therefore, even considering that the problem can be
analytically be solved in its full generality, from now we focus our attention on the behavior of the solutions derived
by taking the limit $\Omega_{\Lambda_{0i}} \rightarrow 0$. In this case, the scale factor given by Eq. (2.11) reduces to 
\begin{equation}\label{a(t)1}
a(t)=a_i\left[\frac{3}{2}(1+\omega)(1-\beta)H_i (t_c -t) \right]^{\frac{2}{3(1+\omega)(1-\beta)}}.
\end{equation}
It is worth mentioning that apart the physical choice of constants, the above solution for $\beta=0$ reduces to the one derived by Cai and Wang  \cite{CaiWang} in their study of a collapsing one  fluid component. From Eq. (2.12) we also  see that the collapsing time is given by:
\begin{equation}
t_c=\frac{2H_{i}^{-1}}{3(1+\omega)(1-\beta)}.
\end{equation}
As it appears, the modulus of the initial collapsing Hubble function, $H_i$, sets the time scale to reach the singularity. { Note that in the limiting case $\beta \rightarrow 1$, the collapsing time $t_c \rightarrow \infty$ and the model is nonsingular (pure de Sitter vacuum). From now on it will be assumed that $0 \leq \beta <1$.} 

The following equivalents forms 
for Eq. (2.12) are also useful:
\begin{equation}
a(t)=a_i\left(1-\frac{t}{t_c}\right)^{\frac{2}{3(1+\omega)(1-\beta)}},
\end{equation}
and
\begin{equation}\label{a(t)1}
a(t)=a_i\left[1-\frac{3}{2}(1+\omega)(1-\beta)H_{i}t \right]^{\frac{2}{3(1+\omega)(1-\beta)}}.
\end{equation}

\begin{figure*}[th]
			\centerline{\psfig{figure=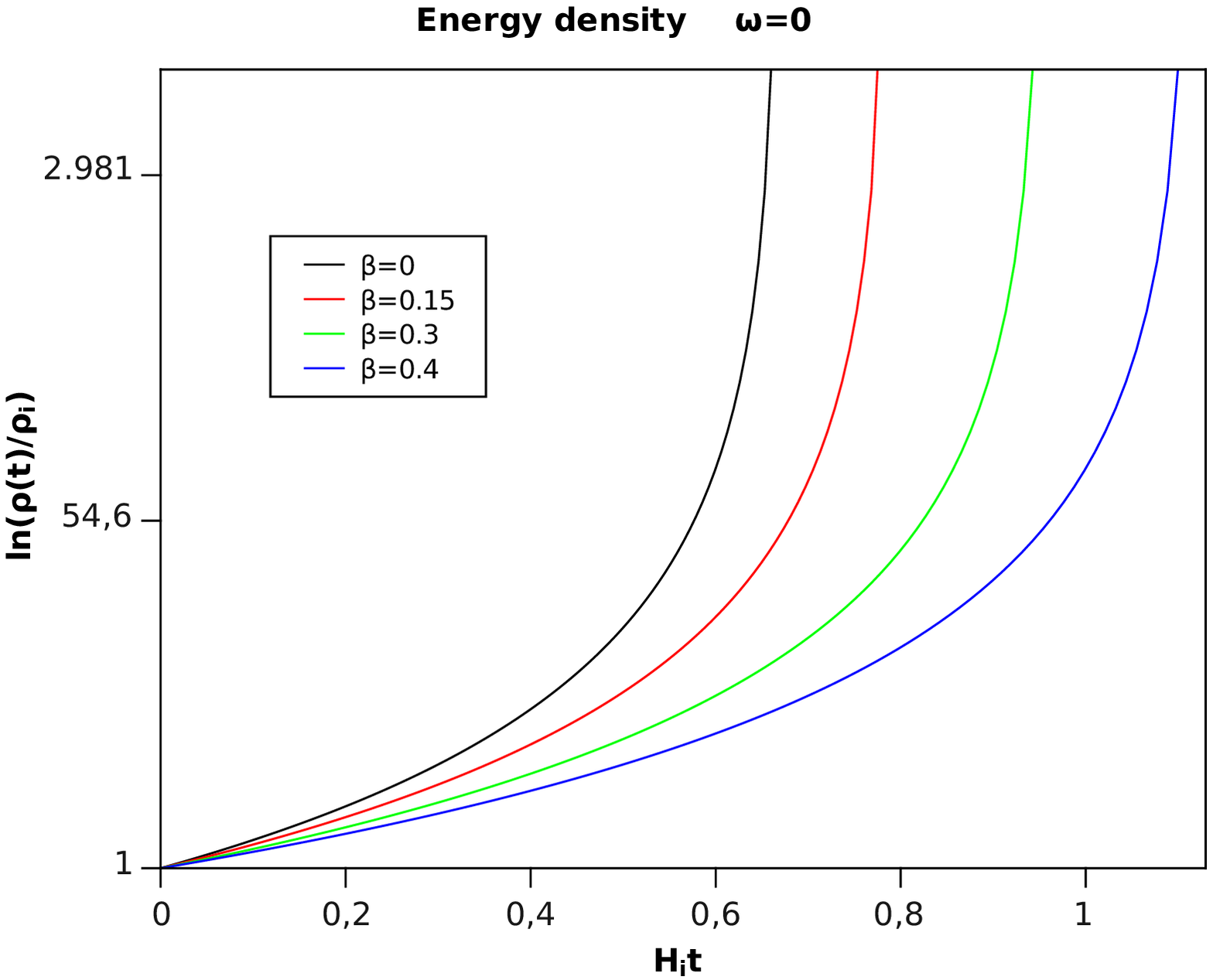,width=3.0truein,height=1.9truein}
			  \psfig{figure=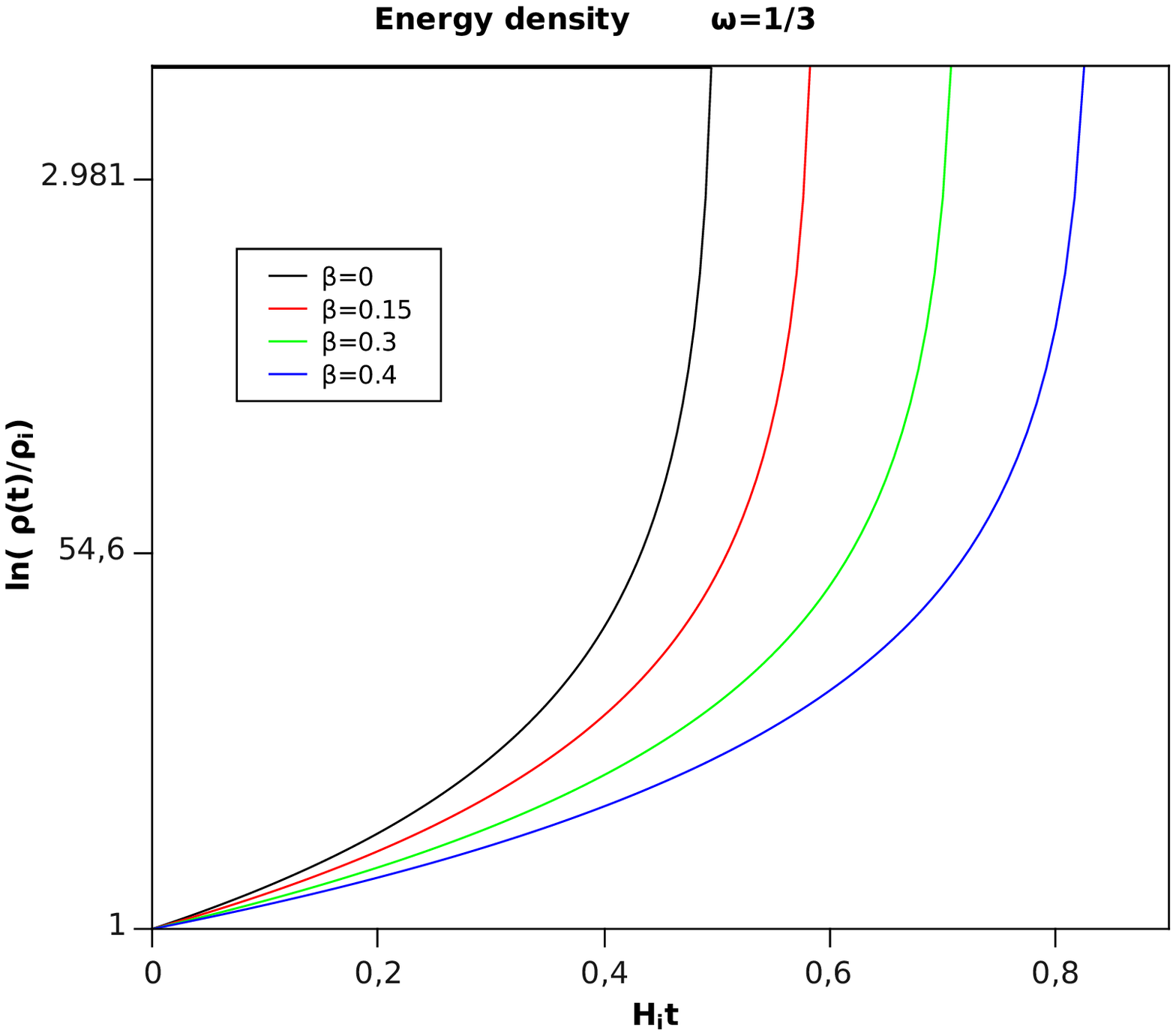,width=3.0truein,height=1.9truein}}
  
                    \caption{Evolution of the total energy density. 
                In the left panel we display the total density for a mixture of dust ($\omega=0$) plus a growing vacuum 
                component as a function of the dimensionless time.  In the right panel we show the same plot but 
                now for a radiation fluid ($\omega = 1/3$) plus vacuum. For both cases the evolution is heavily dependent on the values of the $\beta$ parameter.} 
               \label{B}
              \end{figure*}

Note also that the condition $\dot{a}(t)<0$ 
characterizing the the collapse process has been taken into account in the present formulation since the collapsing ``Hubble function" reads: 
\begin{equation}
H(t)=\frac{-H_i}{1-\frac{3}{2}(1+\omega)(1-\beta)H_i t}\, ,
\end{equation}
so that  $H (0)=-H_i$ as should be expected (see comment below Eq. (\ref{a(t)})).

Without loss of generality, from now on we consider the initial scale factor $a_i=1$. By neglecting $\Lambda_0$, Eq. (2.10) describing the acceleration reads:
\begin{equation}
\frac{\ddot a}{a} = \left[1-\frac{3}{2}(1+\omega)(1-\beta) \right]\frac{\dot a ^2}{a^2}\,.
\end{equation}
We see that in the limiting case, $\beta=1/3$ and $\omega =0$ (dust case), we have $\ddot a=0$; whereas for $\beta > 1/3$ 
the gravitational collapsing process happens in an accelerating way ($\ddot a > 0$). On the other hand, considering $\beta =1/2$ and $\omega =1/3$ 
(radiation case), we have an identical behavior. This is the first vacuum effect.  Only for $\beta = 0$ the standard result is recovered (see \cite{CaiWang}).

In Figure 1,  we display the behavior of the scale factor as a function of the dimensionless time, $T=H_it$, and some 
selected values of the vacuum $\beta$-parameter. Two different scenarios are considered: (i) a dust dust-filled core ($\omega=0$) coupled to 
a growing vacuum component (left), and (ii) radiation ($\omega=1/3$) plus a growing vacuum (right). 
Note that for fixed values of $H_i$  and $\omega$, the collapsing time grows for greater values of $\beta$. This is the second vacuum effect. 

As above mentioned, in the case of dust ($\omega =0$) the evolution of the scale factor 
is altered when $\beta = 1/3$ (see green line in the left panel of Fig. 1). Similarly, for $\omega = 1/3$, the scale factor modify 
its evolution at $\beta=1/2$ (blue line in the right panel of Fig. 1).

To close this section of exact results we write below the energy density for the vacuum and fluid components: 
\begin{eqnarray}
\rho _v (t) &=&\frac{3 \beta \rho _{i}}{\left[1-\frac{3}{2}(1+\omega)(1-\beta)H_i t\right]^2}\, , \\
\rho _f (t) &=& \frac{3(1-\beta) \rho _{i}}{\left[1-\frac{3}{2}(1+\omega)(1-\beta)H_i t\right]^2}\, .
\end{eqnarray}
where $\rho _{i} = H_{i}^2/{8\pi G}$. 

In Figure \ref{B}, we show the time behavior for the total energy density. 
For all values of  $\beta < 1$, we see that the energy density diverges at the collapse time ($t_c$) which is strongly dependent on the values of $\beta$. 
As should be expected, for a given value of $\beta$, the  collapsing time for a coupled radiation component ($\omega = 1/3$) is also reduced in comparison to the dust case.  In a more realistic treatment, a transition from radiation ($\omega=1/3$) to the limit case described by the Zeldovich's stiff-matter medium  ($\omega =1$) may occur at the late stages. Naturally, such a final state is also included in the general solutions for $a(t)$ and $\rho(t)$ with similar plots appearing in Figs. 1 and 2.  
\section{Apparent horizon and collapse}
In the FRW spacetime, the observers describing the behavior of the matter fields are comoving with the fluid volume elements. This means that we can 
define a constant geometrical radius for the surface dividing the star interior from the exterior, 
namely $r_\Sigma $.  For such surface,  the metric can be written as:
\begin{equation}
ds^2 _\Sigma = d\tau ^2 -R\left(\tau\right) ^2 d\Omega ^2 \, ,
\end{equation}
where $t=\tau $ and $R(t) = r_\Sigma a(\tau)$.

The decision about the  final stage of the collapse process is closely related to  the emerging  apparent horizon which must be 
formed  before the collapsing time ($t=t_c$), that is, when the real singular point is attained. Apparent horizons are space-like surfaces
with future point converging null geodesics on both sides of the surface \cite{Hawking}. For an initially untrapped star, when the apparent horizon
appears before the singularity, one may say that a BH is formed, otherwise a
naked singularity is the final stage of the collapse.

The formation of the apparent horizon is driven by the condition \cite{McV,CM73,CaiWang}:
\begin{equation}
R_{,\alpha} R_{,\beta}g^{\alpha \beta}= \left( r \dot a \right) ^2 -1 =0 \, ,
\end{equation}
where $()_{,x}=\frac{\partial }{\partial x}$ and $R(t,r)=ra(t)$.

\begin{figure*}[th]
                 \centerline{\psfig{figure=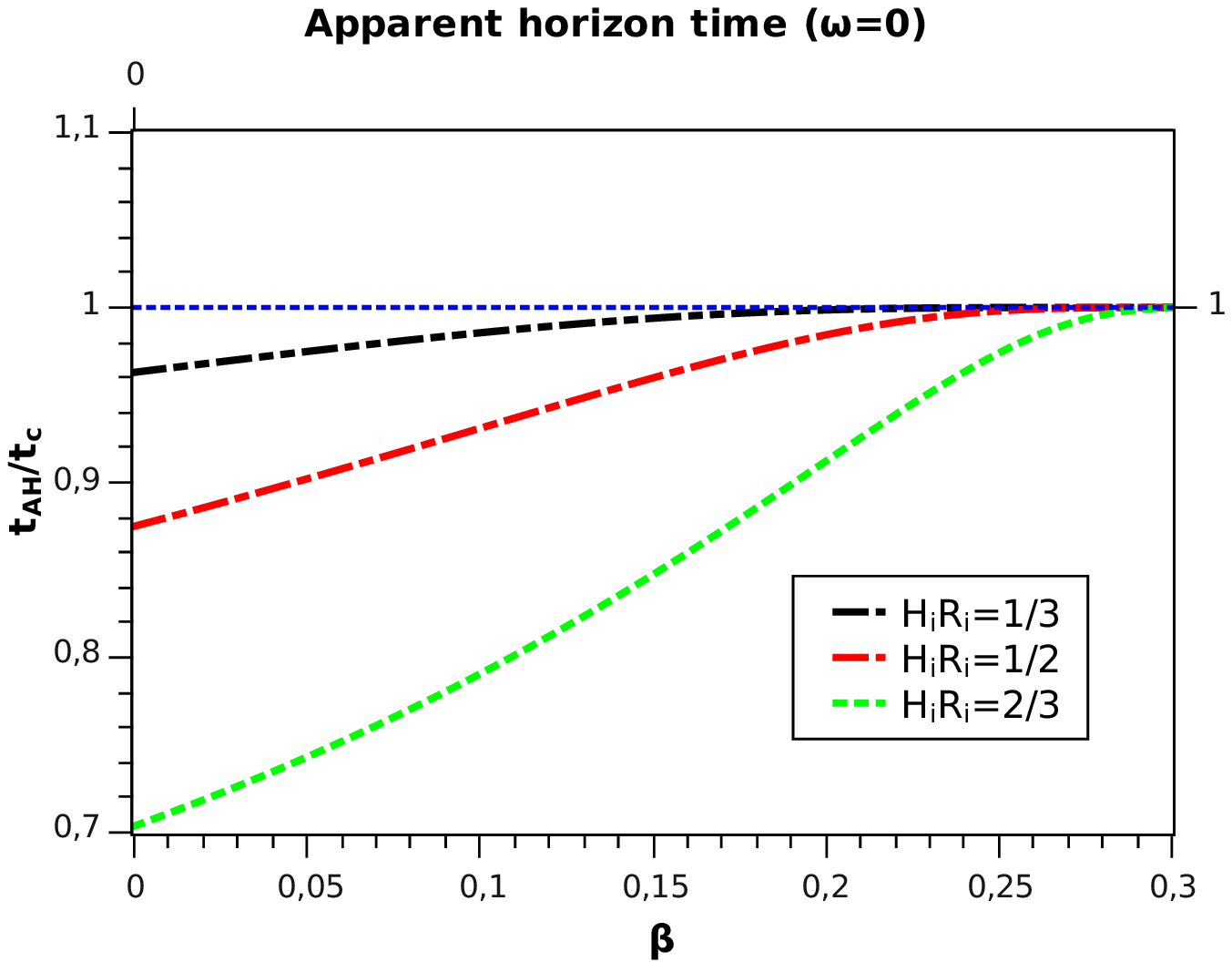,width=3.0truein,height=1.9truein}
                    \psfig{figure=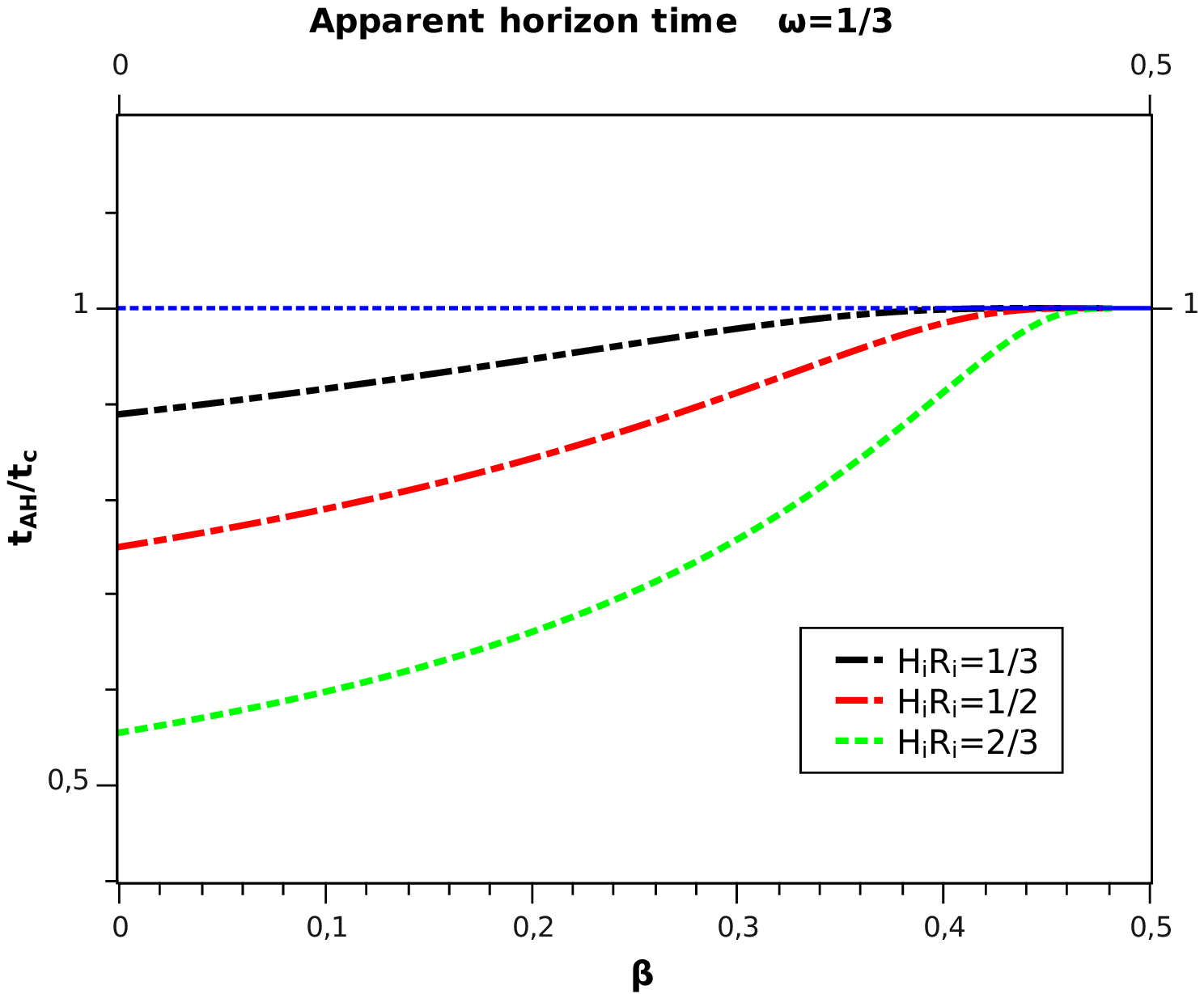,width=3.0truein,height=1.9truein}}
                 \caption{ { The dimensionless ratio $t_{AH}/t_c$ as a function of the $\beta$ parameter (see Eq. 2.26). As in the previous figure, we display the behavior for dust (left panel) and radiation (right panel) cases. For a fixed initial condition ($H_iR_i$),  the formation of the horizon is delayed for higher values of $\beta$ as long as condition (2.27) is satisfied. When  (2.27) becomes an equality, that is, for $\beta=1/3$ \,(dust) and $1/2$ \,(radiation)  we see that $t_{AH}=t_c$ and the corresponding effective mass goes to zero (see Fig. 4). For greater values of $\beta$, the effective mass becomes  negative and the corresponding singularities are naked.}}
                \label{C}
               \end{figure*} 

In this work, we assume that the star is initially not trapped, and, as such, the comoving surface is spacelike. Hence   
\begin{equation}
R_{,\alpha}R_{,\beta}g^{\alpha \beta}=\left[ r_\Sigma \dot{a}\left( t_i \right) \right] ^2 -1 < 0\, ,
\end{equation}
which implies that $0<R_iH_i<1$. Such a domain for the product $R_i H_i$ will be useful when we discuss ahead the criteria for BH formation.
Likewise, other important physical quantity is the mass function that furnish the total mass inside  the surface with radius $r$ at time $t$. 
Originally, Cahill and McVittie \cite{CM73} wrote such a function for a particular reference system that here takes the following form \cite{CaiWang}
\begin{equation}
m(t,r)=\frac{1}{2}R\left( 1+R_{,\alpha}R_{,\beta}g^{\alpha \beta}\right) =\frac{1}{2}R \dot{R}^2 \, ,
\end{equation}
that appear in the literature more frequently \cite{Poisson}.
           
In our study, the condition to bring into being the apparent horizon assumes the form
\begin{equation}
\dot{R}=R_i H_i \left[ 1-\frac{3}{2}(1+\omega)(1-\beta)H_i t_{AH}\right]^{\frac{2-3(1+\omega)(1-\beta)}{3(1+\omega)(1-\beta)}}=1\, ,
\end{equation}
where $t_{AH}$ is the time marking the apparent horizon formation:
\begin{equation}
t_{AH}=\frac{2{H_i}^{-1}}{3(1+\omega)(1-\beta)}\left[1-(R_i H_i)^{\frac{3(1+\omega)(1-\beta)}{3(1+\omega)(1-\beta)-2}} \right]\, ,
\end{equation}
or equivalently (see Eq. (2.13))
\begin{equation}
\frac{t_{AH}}{t_c} =  \left[1-(R_i H_i)^{\frac{3(1+\omega)(1-\beta)}{3(1+\omega)(1-\beta)-2}} \right]\, .
\end{equation}

{ In particular, for $\omega=\beta = 0$,  $H_i \sim 30sec^{-1}$, 
and $H_iR_i \sim 1/2$ we find $t_{AH} \sim 2\times10^{-2}sec$, while for $\omega=0,\, \beta = 0.2$ we obtain $t_{AH} \sim 3\times10^{-2}sec$.
The formation time of the apparent horizon is strongly correlated with $\beta$.

In Figure \ref{C}, we display  the behavior of  the dimensionless ratio, ${t_{AH}}/{t_c}$,  as a function of the $\beta$
parameter for some selected values of the product $R_i H_i$.}  Since the product $R_i H_i$ is positive and smaller 
than unity (see discussion below Eq. (2.22)), as long as the quotient in the exponent of Eq. (2.26) is greater than zero, 
the formation of the apparent horizon will occur before the collapsing time $t_c$ ($t_{AH} < t_c$). As one may check,  this condition is defined by 

\begin{equation}
\beta < 1-\frac{2}{3(1+w)}\, .
\end{equation}
In the present context, the above relation describes a kind of compromise between the dynamics of the collapsing system and
 the formation of the apparent horizon. Naturally, when such a condition is violated within the accelerating collapsing mixture (fluid plus vacuum),
 the formation of the apparent horizon it will be  avoided thereby giving rise to a  pure naked singularity.    

{ At this point one may ask about the specific signatures of  black holes and naked singularities. 
In other words, how to discriminate observationally these two singular structures? As far as we know, 
there is no experimental suggestion aiming to distinguish them based uniquely on the physics of the collapsing process.
  It is also not clear whether the very energetic events named gamma-ray bursts (GRB's) are somewhat related to the
 formation of spacetime singularities. Some aspects of gravitational collapse and spacetime singularities containing
 a discussion about such a possibility was recently published by Joshi and Malafarina \cite{Malafarina}. Nevertheless,
 some authors have claimed  that observations involving strong gravitational lensing \cite{Virbahadra,Sahu} 
and accretion disks \cite{Kovac} are able to discriminate black holes from naked singularities. 

In the strong lensing regime, for instance, it was found that the number of relativistic images and Einstein rings formed in the case of naked singularities are more separated from each other than in the case of black holes \cite{Sahu}. More recently, Kov\'acs and Harko \cite{Kovac} also argued that the thermodynamic and electromagnetic features of accretion disks are different for this two classes of objects thereby giving a clear cut signature that could distinguish  such spacetime singularities. In particular, they show that the conversion efficiency of the accretion mass into radiation to the case of rotating naked  singularities  is always higher than that of black holes (see their table I). 

On the other hand, there are some useful constraints on the $\beta$ parameter given by observations coming from the
cosmic dark sector. For instance, Birkel and Sarkar \cite{Birkel} derived the upper limit $\beta < 0.13$  by using big-bang nucleosynthesis
(BBN) in the presence of a decaying vacuum. Later on, Lima et al. \cite{ALima} rediscussed this bound  inferred from BBN 
thereby obtaining $\beta \leq 0.16$. It is also known that a possible effect related  to an adiabatic decaying vacuum into 
thermalized photons  is to alter the redshift temperature law of CMB to $T = T_0 (1 + z)^{1 - \beta}$ \cite{Lima1996}. 
In this case, several authors have constrained  $\beta$  by using different cosmological probes \cite{probes}.  
Such results were obtained by assuming that the decaying vacuum is coupled only with photons but other decay channels 
are not forbidden from first principles.  For instance, in the 
context of interacting dark energy models it may decay into dark matter ($\omega=0$). In this case, Basilakos    
\cite{Basilakos2}  obtained  $\beta \sim 0.004$ ($\gamma$ in his notation) for a coupling with dust.
All the results  above mentioned are in agreement with our condition (2.27), however, higher values of $\beta$ for
 collapsing systems are not forbidden from first principles.}

\begin{figure*}[th]
\centerline{\psfig{figure=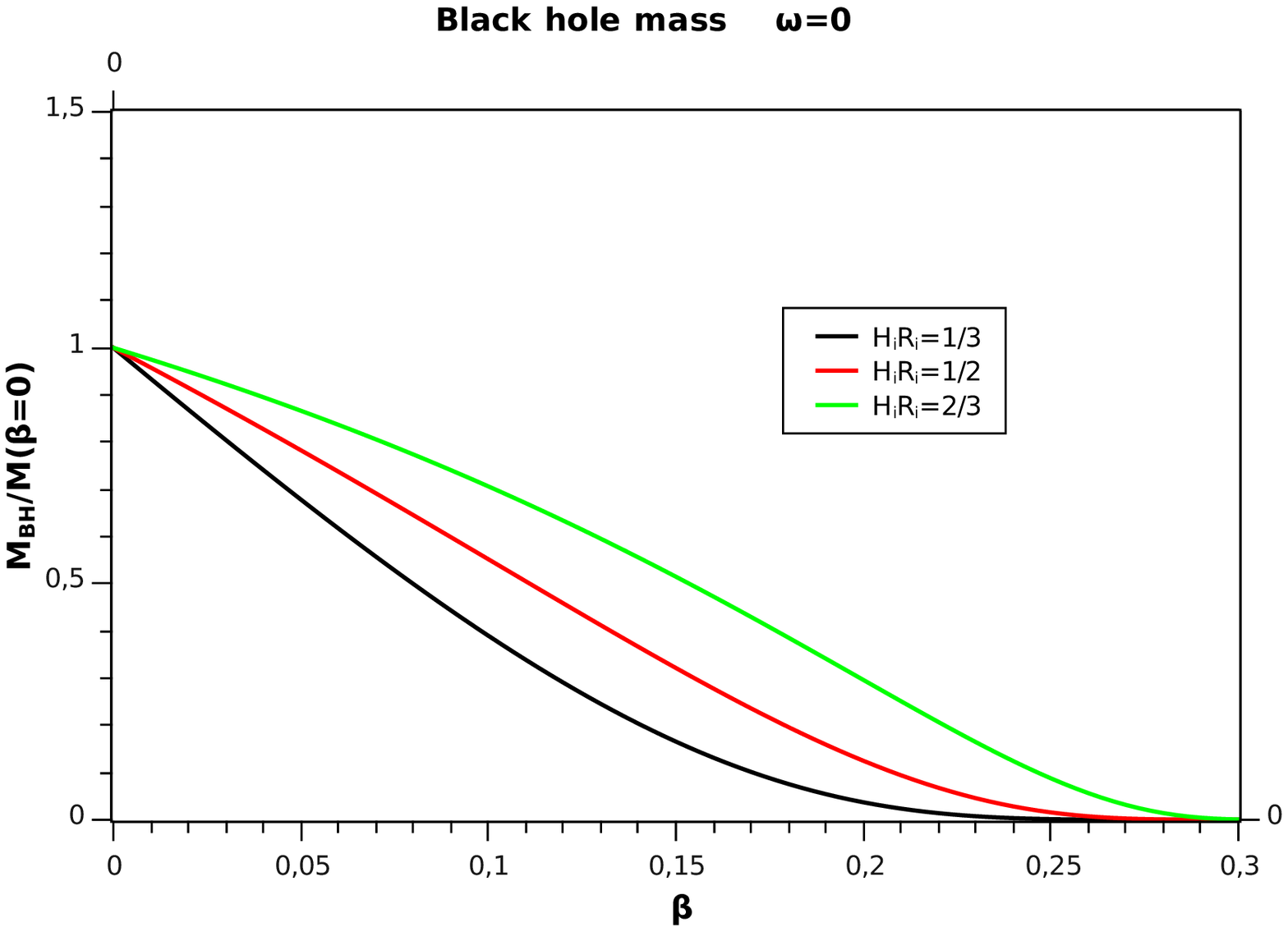,width=3.0truein,height=1.9truein}
\psfig{figure=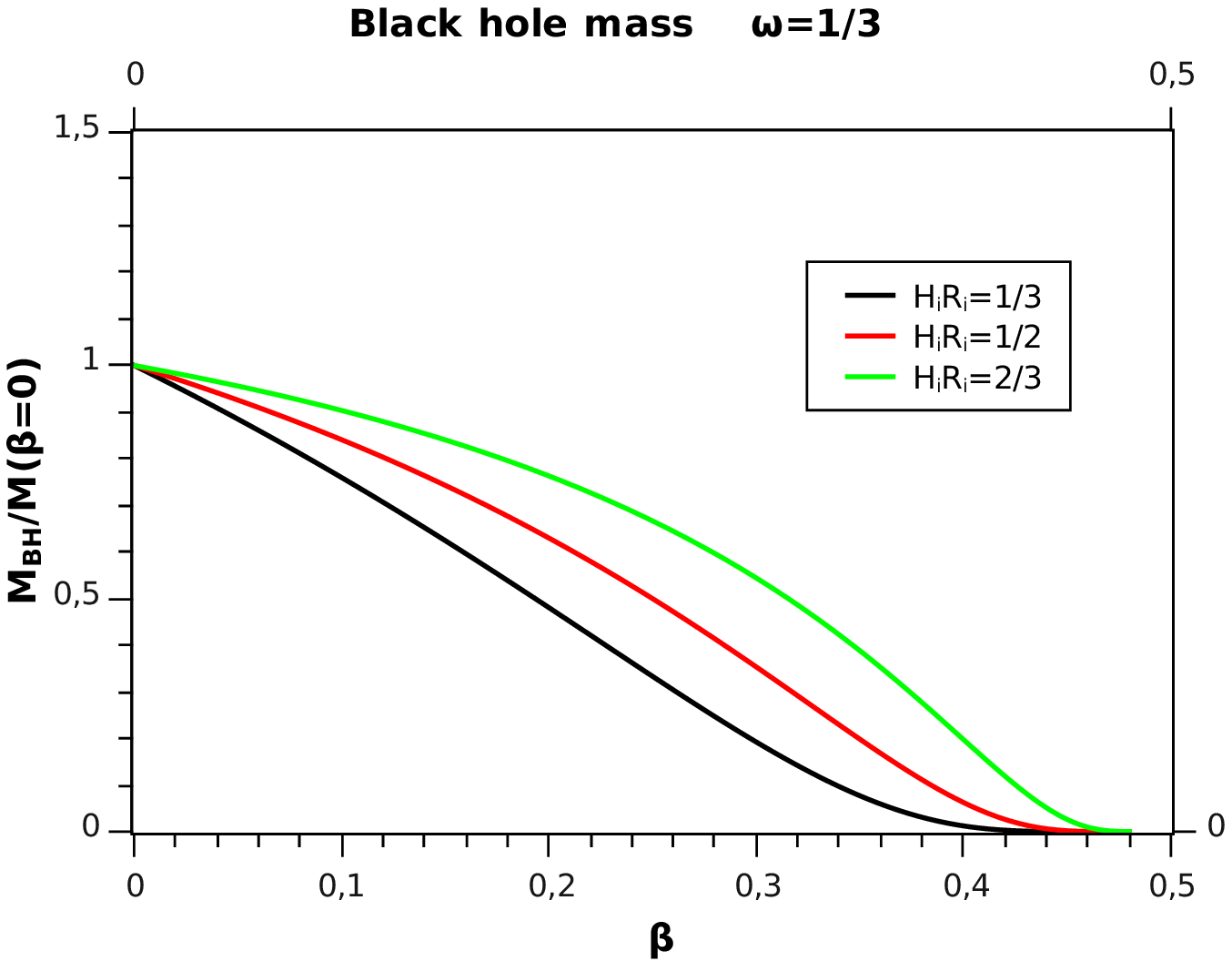,width=3.0truein,height=1.9truein}}
                                
\caption{{ The ratio between the mass of the BH formed and the mass of a pure fluid versus the $\beta$ parameter.}  
As in the other plots we also consider separately the dust case plus the interacting vacuum (left panel), 
and the radiation case plus vacuum (right panel).  For a fixed value of the product $H_iR_i$, we see that  
the BH mass is not only heavily dependent on the $\beta$ parameter but decreases for higher values of
 $\beta$ satisfying the constraint given by Eq. (2.27). All these results are based on the Cahill and McVittie mass definition \cite{CM73}.}
\label{D}
\end{figure*} 

Now, in order to complete our description let us discuss the collapsed mass inside the radius $r$ at time $t$ (see Eq. (2.23)). It can be written as:
\begin{equation}
m(r,t)=\frac{1}{2}R_i^3H_i ^2\left\lbrace 1-\frac{3}{2}(1+\omega)(1-\beta)H_it \right\rbrace ^{\frac{2\left[\beta(1+\omega)-\omega \right]}{(1+\omega)(1-\beta)}}.
\end{equation}
This quantity yields the total mass of the collapsing star at time $\tau$ within the surface $r_\Sigma$,  that is,
  $M(\tau) = m(r_\Sigma, \tau)$. With the help of Eqs. (2.25) and (2.28) we find the expression for the total
 collapsed mass inside the apparent horizon, namely:
\begin{equation}
M(\tau _{AH})=\frac{1}{2}R_i^3H_i^2(R_i H_i)^{\frac{3(1+\omega)(1-\beta)-3}{1-\frac{3}{2}(1+\omega)(1-\beta)}} \, ,
\end{equation}
which reduces to $M(\tau_{AH})= \frac{1}{2}R_i^3H_i^2$ in the case of a pure dust fluid ($\omega=\beta=0$).
{ In physical units, the above mass can be rewritten as
\begin{equation}
M(\tau _{AH})= \frac{R_i M_{\odot}}{r_{S\odot}} (R_i H_i)^{\frac{2}{{{3}(1+\omega)(1-\beta)} - 2}} \, ,
\end{equation}
where $M_{\odot}$ and $r_{S\odot}$ are respectively, the solar mass and the Schwarzschild radius of the sun.
In the case of a pure radiation fluid ($\omega = 1/3$, $\beta=0$), assuming $R_i = 10\, Km$ and $R_iH_i \sim 1/2$  we find that
$M \sim 13.3 \, M_{\odot}$, while for $\beta = 0.25$ and the remaining quantities as given above, a smaller total mass is obtained, 
 $M \sim 6.6 \, M_{\odot}$. As expected, the total mass depends explicitly on the decaying vacuum $\beta$ parameter which 
is constrained by (2.27) when a BH is formed.} 

{ In  Figure \ref{D}, we  show the decaying vacuum effect on the total BH mass. For each fluid component, we 
have plotted the ratio $M_{BH}/M(\beta=0)$ as a function of the $\beta$ parameter.}
For fixed initial conditions, we see that the BH mass is indeed heavily dependent on the $\beta$ parameter. Interestingly, when $\beta =  1-{2}/3(1+w)$,
the total mass of the BH goes to zero. For each kind of fluid ($\omega$) this critical value  of $\beta$,  is the lower limit of the decaying vacuum parameter signalizing the formation of a naked singularity.  The plots for the formation of the apparent horizon (Fig. 3) and mass ratio of 
the formed BHs suggest that the growing  of the repulsive gravitational energy of the coupled vacuum alters considerably  
the collapse process. As remarked before, for all cases violating the condition given by Eq. (2.27),  the collapsing time $t_c$  remains finite but $t_{AH} > t_c$, and,  as such, one must conclude again that the spacetime singularity is naked (for the fixed initial conditions).

\section{Final Comments}

In this paper we have studied the gravitational collapse of a spherically symmetric
star with finite radius filled by a homogeneous and isotropic fluid obeying the EoS, $p=\omega \rho$, plus an interacting 
growing vacuum energy density. We stress that all the curvature effects were neglected and that the dynamical $\Lambda(t)$-term
  was assumed to obey the relation \cite{CLW,SS}: $\Lambda$ = $\Lambda_0$ + 3$\beta H^{2}$. 

In our approach, all physical constants appearing in the solutions were properly identified. After highlighting
some of their consequences,  we have focused our study on the influence of a growing vacuum energy density on the
BH mass enclosed by the trapped surface and the formation of a naked singularity.  In the applications only 
positive values of the EoS parameter ($\omega$) were studied with special attention to dust ($\omega=0$) and radiation ($\omega=1/3$).
However, the general solutions hold even for negative values of the $\omega$-parameter. Therefore, we have described a two fluid collapsing mixture whose nature is characterized by a two-parametric ($\omega$, $\beta$) phenomenological approach.  

It was found that the final stages of the gravitational collapse depend heavily on the values assumed by  the $\beta$-parameter. For any positive value of the $\omega$ parameter, the collapsing process is delayed by the vacuum energy component. However, the condition constraining the formation of the event horizon changes for distinct  values of $\omega$ (see Eq. (2.27)). In particular, for a dust fluid the condition is  $\beta < \frac{1}{3}$ while for radiation the formation of a BH is allowed for  $\beta < \frac{1}{2}$. When such conditions are not obeyed a naked singularity is formed. {  The critical value, $\beta = 1 - 2/3(1 + \omega)$, defining the boundary between black holes and naked singularities implies that the total mass is zero (see Fig. 4). It should be stressed, however, that such an effect is not related to any dynamical mass evaporation process, like  Hawking radiation. It is closely related with the vacuum pressure, and, generically, must appear when  sufficiently large negative pressures takes place in the matter content.}   

Naturally, all the results derived here are heavily dependent on the  form assumed for $\Lambda(t)$ and even the avoidance of the singularity may occur whether a more realistic description of the decaying vacuum is adopted. Nonetheless, the results obtained
here suggest that a growing vacuum energy density may lead at the late stages to naked singularities even when inhomogeneities are taken into account. 
The physical  effects of a growing time varying vacuum  energy density on the inhomogeneous collapse will be discussed in a forthcoming communication.  

Finally, it is also clear that the hypothesis of  homogeneity and isotropy of the spacetime  of 
the collapsing star is the main caveat of the  proposed model since the pressure should be made to vanish at the boundary
thereby obtaining a smooth matching with the Schwarzschild-de Sitter vacuum solution outside the star. 
\section{Acknowledgments}
The authors are grateful to an anonymous referee whose questions and comments contributed to improve the manuscript.   
M.C. is partially supported by a grant from CNPq and J.A.S.L. is partially supported by  CNPq and FAPESP (No. 04/13668-0).


\begin{thebibliography}{100}

\bibitem{Riess} A. Riess {\it et al.}, Astron. J. {\bf{116}}, 1009
(1998); S. Perlmutter {\it et al.}, Nature, {\bf{ 391}}, 51 (1998); M. Kowalski {\it et al.}, Astrophys. J. {\bf{686}}, 749 (2008); R. Amanullah {\it et al.}, Astrophys. J. {\bf{716}}, 712 (2010). 

\bibitem{Komatsu} D.~N.~Spergel {\it et al.}, {Astrophys.\ J.\ Suppl.} {\bf 148}, 175 (2003). D. N. Spergel et al. Astrophys. J. Suppl. Ser. {\bf 170}, 377 (2007); E. Komatsu {\it et al.}  Astrophys. J. {\bf 192}, 18 (2011). 

\bibitem{rev1} P. J. E. Peebles and B. Ratra, Rev.~Mod.~Phys.
{\bf 75} 559 (2003); T. Padmanabhan, Phys.~Rept. {\bf
380}, 235 (2003); J.~A.~S. Lima, Braz.~Journ.~Phys.
{\bf 34}, 194 (2004), astro-ph/0402109; E. J. M. Copeland and ~S. Tsujikawa,  
Int.~J.~Mod.~Phys. {\bf D15}, 1753 (2006); J. A. Frieman, M. S. Turner and D. Huterer,  
Ann.~Rev.~Astron. \& Astrophys. {\bf 46}, 385 (2008); M. Li {\it et al.},  arXiv:1103.5870 (2011).

\bibitem {Bas2010} S. Basilakos and J. A. S. Lima, Phys. Rev. {\bf D82}, 023504 (2010), [arXiv:1003.5754]; 
S. Basilakos, M. Plionis and J. A. S. Lima Phys. Rev. {\bf D82}, 083517 (2010), arXiv:1103.1464 [astro-ph.CO]. 

\bibitem{Weinberg} S. Weinberg, {Rev. Mod. Phys.} {\bf 61}, 1 (1989).

\bibitem{B33} M. Bronstein, Phys. Z. Sowjetunion {\bf 3}, (1933).

\bibitem{OT86} M. Ozer and M. O. Taha, Phys. Lett.  {\bf B171}, 363 (1986); Nucl.
Phys. B {\bf 287} 776 (1987).

\bibitem{L1} K. Freese et al., {Nucl. Phys. }, {\bf B287}, 797 (1987);  
W. Chen and Y-S. Wu, {Phys. Rev.} {\bf D41}, 695 (1990); D.
Pavo´n, {Phys. Rev. } {\bf D43}, 375 (1991).

\bibitem{CLW} J. C. Carvalho, J. A. S. Lima, and I. Waga, { Phys. Rev. } {\bf D46}, 2404
(1992).

\bibitem {L2} J. A. S. Lima and J. M. F. Maia, {Phys. Rev. } {\bf D49},
5597 (1994);  J. A. S. Lima and M. Trodden, Phys. Rev. {\bf D53}, 4280 (1996), 
[astro-ph/9508049].

\bibitem{L3}  A. I. Arbab and A. M. M. Abdel-Rahman,
{Phys. Rev. } {\bf D50}, 7725 (1994); J. M. Overduin and F. I.
Cooperstock, Phys. Rev.  {\bf D58}, 043506 (1998); J. M.
Overduin, Astrophys. J. {\bf 517}, L1 (1999); M. V. John and
K. B. Joseph, {\it Phys. Rev.} {\bf D61}, 087304 (2000); O.
Bertolami and P. J. Martins, Phys. Rev. {\bf D61}, 064007
(2000);  R. G. Vishwakarma,  Gen. Relativ. Gravit. {\bf 33},
1973 (2001); A. S. Al-Rawaf, {Mod. Phys. Lett.} {\bf A16},
633 ( 2001); M. K. Mak, J. A. Belinchon, and T. Harko,
{Int. J. Mod. Phys.} {\bf D14}, 1265 (2002); M. R. Mbonye, Int.
J. Mod. Phys.  {\bf A18}, 811 (2003); J. V. Cunha and R. C.
Santos, Int. J. Mod. Phys.  {\bf D13}, 1321 (2004); S. Carneiro
and J. A. S. Lima, Int. J. Mod. Phys.  {\bf A20}, 2465 (2005), gr-qc/0405141.

\bibitem{L4} I. Waga, Astrophys. J. {\bf 414}, 436 (1993); L. F. Bloomfield
Torres and I. Waga, Mon. Not. R. Astron. Soc. {\bf 279}, 712
(1996); R. G. Vishwakarma, Class. Quant. Grav. {\bf 17}, 3833
(2000); R. G. Vishwakarma, Class. Quant. Grav. {\bf 18}, 1159
(2001).

\bibitem{ML02} J. M. F. Maia and J. A. S. Lima, Phys. Rev. {\bf D65}, 083513 (2002), arXiv:astro-ph/0112091.

\bibitem{EA1} F.~E.~M.~Costa, J.~S.~Alcaniz and J.~M.~F.~Maia, Phys.\ Rev.\  D {\bf 77}, 083516 (2008). 

\bibitem{Lag1} N. J. Poplawski, [arXiv:gr-qc/0608031v2] (2006); 
T. Harko,  F. S. N. Lobo,  Shin'ichi Nojiri and Sergei D. Odintsov, Phys. Rev. D84 (2011) 024020. 

\bibitem{Penrose}
R. Penrose, {\it Nuovo Cimento Soc. Ital. Fis.} {\bf 1} , 252, (1969).

\bibitem{Papa} A. Papapetrou, {\it A Random Walk in Relativity and Cosmology}, N. Dadhich, J. K. Rao, J. V. Narlikar, 
and C. V. Vishveshwara. Eds. Jonh Wiley 
\& Sons, New York, pp 184-191, (1985).

 \bibitem{Collapse} P. S. Joshi, {\it Global Aspects in Gravitation
      and Cosmology}, Clarendon, Oxford, (1993). D. Christodoulou, Ann. Math. {\bf 140}, 607 (1994). For more recent reviews,
      see, e.g., R. Penrose, in {\em Black Holes and Relativistic
      Stars}, edited by R. M. Wald (University of Chicago Press, 1998);
      A. Krolak, Prog. Theor. Phys. Suppl. {\bf 136}, 45 (1999); P. S.
      Joshi, Pramana {\bf 55}, 529 (2000), and P.S. Joshi, ``{\em Cosmic
      Censorship: A Current Perspective}," {\tt gr-qc/0206087} (2002);
      {\em Gravitational Collapse End States}, {\tt gr-qc/0412082}
      (2004), and references therein. A. Beesham, Pramana {\bf 77}, 429 (2011).

\bibitem{Oppenheimer} J. R. Oppenheimer and H. Synyder, Phys. Rev. {\bf 55},
455 (1939).

\bibitem{LL} L. Landau and E. M. Lifschitz, {\it The Classical theory of Fields}, Pergammon Press (1985); 

P. J. E. Peebles, {\it Principles of Cosmology}, Princeton UP (1993).

\bibitem{CaiWang} R-G. Cai and A. Wang, Phys. Rev.  {\bf D73}, 063005 (2006).

\bibitem{Hawking1} S. W. Hawking, in {\it Black Holes}, edited by C. DeWitt and B. S. DeWitt (Gordon and Breach, New York, 1973).


\bibitem{Anninos} P. Anninos {\it et al.} Phys. Rev. {\bf D50}  3801 (1994).

\bibitem{Mortlock}
D. J. Mortlock {\it et al.},  Nature, {\bf 474}, 616, (2011).

\bibitem{SS} I. L. Shapiro and  J. Sol\`a,  JHEP 0202 (2002) 006, hep-th/0012227. For a recent review see J. Sol\`a, J. Phys. Conf. Ser.
283 (2011) 012033, arXiv:1102.1815 [astro-ph.CO]. 

\bibitem{Hawking} S. W. Hawking and G. F. R. Ellis, {\em The Large Scale
Structure of Spacetime}, Cambridge University Press, Cambridge  (1973).

\bibitem{McV} G. C. McVittie, Mon.  Not. R. Astron. Soc. {\bf 93}, 325 (1933).


\bibitem{CM73}  M. E. Cahill and G. C. McVittie, J. Math. Phys. {\bf 11}, 1382 (1970).


\bibitem{Poisson} E. Poisson and W. Israel, Phys. Rev. {\bf D41}  1796 (1990); A. Wang, J. F. Villas da Rocha, and N. O. Santos, 
{\it ibidem} {\bf D56}, 7692 (1997); J. F. Villas da Rocha, A. Wang, and N. O. Santos, Phys. Lett. 
{\bf A255}, 213 (1999); S. A. Hayward, Phys. Rev. {\bf D70}, 104027 (2004); Phys. Rev. Lett. {\bf 93}, 251101 (2004). 

\bibitem{Malafarina}
P. S. Joshi and D. Malafarina, Int. J. Mod. Phys. {\bf D20}, 2641 (2011).


\bibitem{Virbahadra}
K. S. Virbahadra and G. F. R. Ellis, Phys. Rev. {\bf D65}, 103004 (2002).

\bibitem{Sahu}
S. Sahu, M. Patil, D. Narasimha and P. S. Joshi, {\tt arXiv:1206.3077v1} (2012).

\bibitem{Kovac}
Z. Kov\'acs and T. Harko, Phys. Rev. {\bf D82}, 124047 (2010).

\bibitem{Birkel}
M. Birkel and S. Sarkar, Astropart. Phys. {\bf 6}, 197 (1997). 

\bibitem{ALima}
J. A. S. Lima, J. M. F. Maia and N. Pires, IAU Symposium {\bf 198}, 111 (2000).

\bibitem {Lima1996} J. A. S. Lima, Phys. Rev. D {\bf 54}, 2571 (1996),  gr-qc/9605055; 
{\bf ibdem} Gen. Rel. Grav. {\bf 805}, 27 (1997), gr-qc/9605056;  J. A. S. Lima, A. I. Silva and S. M. Viegas, Mon. Not. R. Astron. Soc. {\bf 312}, 747
(2000), astro-ph/9902337.

\bibitem{probes} R. Opher and A. Pelinson, Mon. Not. R. Ast. Soc.  {\bf 362}, 167 (2005); G. Luzzi et al., Astrophys. J. {\bf 705}, 1122 (2009); P. Noterdaeme at, al. Astron. Astrophys. {\bf 526}, L7 (2011).


\bibitem{Basilakos2}
S. Basilakos, Astron. and Astrophys. {\bf 508}, 575 (2009).




\end{thebibliography}
\end{document}